\documentclass[12pt]{iopart}
\usepackage{amssymb,graphicx}
\usepackage{color}
\usepackage[normalem]{ulem}
\usepackage{todonotes}
\usepackage{mcite}
\usepackage{cite}

\newcommand{\dd}{\rmd}

\newcommand{\pd}[2]{\frac{\partial #1}{\partial #2}}

\newcommand{\mean}[1]{\langle #1\rangle}
\newcommand{\Int}[1]{\int\dd #1\;}
\newcommand{\path}[1]{\int\mathcal{D}#1\;}
\newcommand{\IInt}[3]{\int_{#2}^{#3}\dd #1\;}
\newcommand{\bra}[1]{\left\langle #1 \right|}
\newcommand{\ket}[1]{\left| #1 \right\rangle}

\newcommand{\ra}{\rightarrow}
\newcommand{\tobs}{t_\mathrm{obs}}

\newcommand{\rev}[1]{{#1}}
\newcommand{\revd}[1]{}
\newcommand{\revr}[2]{{#2}}

\begin{document}

\title{Meta-work and the analogous Jarzynski relation in ensembles of dynamical trajectories}

\author{Robert M. Turner}
\address{Department of Physics and Astronomy, University of Nottingham, 
  Nottingham, NG7 2RD, United Kingdom}

\author{Thomas Speck}
\address{Institut f\"ur Physik, Johannes Gutenberg-Universit\"at Mainz,
  Staudingerweg 7-9, 55128 Mainz, Germany}

\author{Juan P. Garrahan}
\address{Department of Physics and Astronomy, University of Nottingham, 
  Nottingham, NG7 2RD, United Kingdom}

\begin{abstract}
Recently there has been growing interest in extending the thermodynamic method from static configurations to dynamical trajectories.  In this approach, ensembles of trajectories are treated in an analogous manner to ensembles of configurations in equilibrium statistical mechanics: generating functions of dynamical observables are interpreted as partition sums, and the statistical properties of trajectory ensembles are encoded in free-energy functions that can be obtained through large-deviation methods in a suitable large time limit.  This establishes what one can call a ``thermodynamics of trajectories''.  In this paper we go a step further, and make a first
  connection to fluctuation theorems by generalising them to this dynamical context. 
  We show that an effective ``meta-dynamics'' in
  the space of trajectories gives rise to the celebrated Jarzynski relation
  connecting an appropriately defined ``meta-work'' with changes in dynamical generating functions. We
  demonstrate the potential applicability of this result to computer
  simulations for two open quantum systems, a two-level system and the
  micromaser. We finally discuss the behavior of the Jarzynski relation across
  a first-order trajectory phase transition.
\end{abstract}

\maketitle


\section{Introduction}

The so-called ``thermodynamics of trajectories'' approach provides a
description of time-ordered dynamic events that is analogous to the
thermodynamic description of configurations in space. Using large-deviation
methods~\cite{eckm85,ruel04,touc09}, ensembles of trajectories can be
classified by \emph{dynamic} order parameters and their conjugate fields. This is in effect the thermodynamic formalism of Ruelle~\cite{eckm85,ruel04} applied to the space of trajectories, rather than configurations.
Quantities analogous to free-energy densities and entropy densities have been
identified, and used to gain insight into rare dynamical behaviours of systems
both classical~\cite{lebo99,mero05,leco07,garr07,Baule2008,hedg09,jack10,giar11,Pitard2011,flin13,Chetrite2013,Chikkadi2014,Nemoto2014} and
quantum~\cite{espo09,garr10,budi11,li11,ivan13,gamb12}. Of particular
interest has been the identification of \emph{dynamical phase transitions}
into non-equilibrium states with vastly different dynamic properties. To this
end, the use of transition path sampling (TPS)~\cite{dell02} has allowed for
efficient numerical generation of non-equilibrium states, which has had much
success in describing the dynamics of glassy
systems~\cite{hedg09,elma12,spec12,spec12b}.

Trajectories and their ensembles also play a central role for the theoretical
study of driven non-equilibrium systems that has led to the formulation of a
class of relations, called \emph{fluctuation theorems} \cite{Gallavotti1995,jarz97,jarz97a,Kurchan1998,Crooks1999,Hatano2001,Bustamante2005,Seifert2012}, which hold arbitrarily
far from thermal equilibrium. Of central importance is Jarzynski's
non-equilibrium work relation, which relates the work spent in an arbitrarily
fast switching process to the change of free-energy~\cite{jarz97,jarz97a}.
Given that the thermodynamics of trajectories approach is the generalisation to dynamical ensembles of  equilibrium thermodynamics, it is natural to expect that there will also be an analogous extension 
to trajectory ensembles of the fundamental non-equilibrium relations encoded by the fluctuation theorems.  This is the question we address in this work.

The purpose of this paper is two-fold. First, we introduce the concept of
processes in the space of trajectories resulting from changing the conjugate
field. This allows to identify a meta-work, which, through the analogous
Jarzynski relation, allows the computation of the large deviation
function. Second, we explore this result in computer simulations of two
quantum systems. To this end we employ an algorithm based on transition path
sampling while changing the conjugate field. For computational convenience, we
work with the recently introduced $x$-ensemble, in which the observation time
is the fluctuating order parameter while the number of events is held
fixed~\cite{budi14}. Specifically, we study two open quantum systems, the
dynamics of which is described by Lindblad master equations \cite{gard04}. The first system
we consider is a dissipative two-level system \cite{Plenio1998}\rev{,} which can be easily solved analytically and thus provides \revd{with} a simple illustration of our approach.  
The second model system is the single atom maser, or micromaser~\cite{engl02}. Depending
on the parameters\rev{,} this system exhibits multiple dynamical crossovers, i.e.,
sharp changes of the mean observation time as we change $x$. It thus
allows to investigate the behavior of the Jarzynksi relation as one crosses first-order discontinuities, a situation that seems to have received comparably low
attention (see, e.g., Refs.~\cite{chat06,hart14} for numerical and
Ref.~\cite{impa05a} for a mean-field study in the case of the Ising model).

\section{Thermodynamic formalism and ensembles of trajectories}

The probability distribution of some observable $E$, under rather mild
conditions, \revr{leads to formally the same structure as}{gives rise to a
  formal structure that} is known from thermodynamics. The first condition is
that $E$ is extensive, i.e., there is a ``system size'' $K$ and the mean of
$e=E/K$ remains both nonzero and finite as $K\ra\infty$. Second, the
probability of $E$ takes on a large deviation form, $P(E)\asymp
e^{-K\phi(e)}$, with $\phi(e)$ independent of $K$. For example, in equilibrium
statistical physics, if $K$ is the number of particles in a closed volume and
$E$ the energy, the function $\phi(e)$ is immediately identified as the
negative specific entropy.  In this case, the partition sum
\begin{equation}
  \label{eq:Z:eq}
  Z(\beta) = \Int{E} P(E)e^{-\beta E} \asymp e^{Kg(\beta)}
\end{equation}
also has a large deviation form involving the free-energy per particle
$g(\beta)$, see Ref.~\cite{touc09} for a general introduction. Both entropy
and free-energy are related by a Legendre transform,
\begin{equation}
  \label{eq:lege}
  g(\beta) = -\min_e [\phi(e)+ \beta e],
\end{equation}
which describes the transformation between the micro-canonical ensemble at
fixed $e$ to the canonical ensemble at fixed inverse temperature $\beta$.

This thermodynamic formalism can be extended to dynamic
processes, where now $K$ denotes the the observation time. In this case the
mathematical structure remains the same but one of course loses the immediate
physical interpretation of the canonical ensemble. Here we consider systems
evolving in time due to their physical, stochastic dynamics. Hence, over a
given time $\tobs$ we can define \emph{trajectories}
\begin{equation}
  \chi \equiv \{ z_t | 0 \leqslant t \leqslant \tobs \}
\end{equation}
recording the random sequence of microstates $z$ the system has visited. We
characterize trajectories by an order parameter that plays the role extensive quantities, such as energy, play in conventional thermodynamics.  Examples for these dynamical order parameters
include the total number of transitions (or ``jumps'')~\cite{garr09} in a trajectory, the \revd{to} total number of certain specific events, the time-integral of the 
mobility of particles~\cite{hedg09,spec12}, or the time-integral of the 
number of particles that are part of a specific structure~\cite{spec12b}. 
For clarity of presentation, we consider a single order parameter but the
extension to more than one is straightforward. The crucial property of
admissible order parameters is that they are extensive in space and time.

The parameter $K$\rev{,} which determines the size of trajectories and the
corresponding large-size limit, can be something other than the total
observation time \cite{Bolhuis2008,budi14}. 
In keeping with our thermodynamic analogy, this would correspond to two distinct trajectory ensembles.  
For definiteness, we will work here specifically with the recently introduced
$x$-ensemble~\cite{budi14} although our results are valid more generally.
Consider a system in which it is possible to identify and count some event,
the specific nature of the event is unimportant, and could be, for example,
photon emissions from an atomic system. These events are separated by waiting
times $t_n$. We define the probability that observing $K$ events takes a
particular amount of time, $\tobs\equiv\sum_{n=1}^K t_n$, which in the
large-$K$ limit takes on a large deviation form
\begin{equation}
  \label{eq:P}
  P_K(\tobs) \equiv \path{\chi} \rho_0(\chi)\delta(\tobs(\chi)-\tobs)
  \asymp e^{-K\phi(\tau)},
\end{equation}
where $\rho_0(\chi)$ denotes the probability distribution of trajectories,
as given by the dynamics under consideration, 
$\tau\equiv\tobs/K$ is the \revd{intensive} average waiting time within a single
trajectory, and the rate function $\phi(\tau)$ quantifies how fluctuations of
$\tau$ decay as the number of events is increased. The functional measure
$\mathcal D\chi$ of paths implies normalization, $\path{\chi}\rho_0(\chi)=1$.

Taking the Laplace transform of the probability Eq.~(\ref{eq:P}) defines the
moment generating function
\begin{equation}
  \label{eq:Z}
  Z_K(x) \equiv \IInt{\tobs}{0}{\infty} P_K(\tobs)e^{-x\tobs} \asymp e^{Kg(x)}.
\end{equation}
Its logarithm defines the cumulant generating function (CGF) $g(x,K)\equiv\ln
Z_K(x)/K$, which also has a large-deviation form in the limit of which $g(x)$
becomes independent of $K$. In analogy with thermodynamics, $\phi(\tau)$ and
$g(x)$ are identified as the associated (negative) entropy density and
free-energy density, respectively, which are related through the Legendre
transform Eq.~(\ref{eq:lege}). Pursuing
the analogy with thermodynamics further through identifying the number of
events $K$ with particle numbers and the trajectory length $\tobs$ with a
volume, $x$ is analogous to the field conjugate to volume with fixed particle
numbers, i.e., a pressure. We have thus introduced a ``canonical'' ensemble of
trajectories~\cite{chet13}
\begin{equation}
  \label{eq:path}
  \rho_x(\chi) \equiv \frac{\rho_0(\chi)e^{-x\tobs(\chi)}}{Z_K(x)}
\end{equation}
where $\rho_x(\chi)$ is the probability of a trajectory $\chi$ at fixed $x$ (rather than fixed $K$). Physical dynamics takes place at $x=0$, while $x\neq0$ probes the statistics of atypical trajectories.
For details see \cite{budi14}.

\section{Jarzynski relation in trajectory space}

\subsection{Meta-dynamics: Dynamics in the space of trajectories}

The situation considered by the Jarzynski relation is that of a system
initially in thermal equilibrium with a heat reservoir\revd{s}, where the system is
subsequently driven away from equilibrium by externally changing one or more
parameters. The dynamics of the system obeys detailed balance with respect to
the stationary distribution corresponding to the instantaneous values of the
control parameters. Non-equilibrium can then be described as a ``lag'' between
this stationary distribution and the actual distribution~\cite{vaik09}. The
Jarzynski relation~\cite{jarz97} relates the average over all trajectories
with the change of free-energy between initial and final state.

In the trajectory ensemble, we are interested in a very similar situation,
where we want to determine the function $g(x)$ over a range of values
$x$. Instead of performing many ``equilibrated'' simulation runs at fixed $x$,
we aim to extract the function $g(x)$ while changing $x$. To this end we
require to notion of a meta-dynamics and a meta-time, which for convenience we take as integer, 
enumerating the sequence of
generated trajectories $\vec\chi\equiv(\chi_0,\dots,\chi_N)$. The meta-dynamics
that generates these trajectories is required to obey detailed balance with respect to the distribution $\rho_x(\chi)$, that is, 
\begin{equation}
  \label{eq:db}
  \rho_x(\chi) p_x(\chi\ra\chi') = \rho_x(\chi') p_x(\chi'\ra\chi),
\end{equation}
where $p_x(\chi\ra\chi')$ is the probability to generate the trajectory $\chi$ given a
previous trajectory $\chi'$, and $\rho_x(\chi)$ is defined in
Eq.~(\ref{eq:path}).  Natural candidates for the algorithm used to generate new
trajectories are based on \emph{transition path sampling} and the specific
algorithm used in this work is that of \cite{budi14} (also detailed for completeness in~\ref{sec:algo}).

\subsection{Meta-work and the Jarzynski relation}

Equation (\ref{eq:path}) has the form of an equilibrium Boltzmann distribution,
where $E_x(\chi)=x\tobs(\chi)$ can be identified as the analog of an
``energy''. Suppose that we change $x$ along the sequence $\vec\chi$: We start
with a value $x_0$ for the biasing field and generate the initial trajectory
$\chi_0$. We then change the value of $x_0$ to $x_1$ and generate the next
trajectory $\chi_1$ of the sequence and so on. The change of the ``energy''
along the whole sequence is
\begin{equation}
  \Delta E \equiv E_{x_N}(\chi_N) - E_{x_0}(\chi_0) = W + Q,
\end{equation}
which can be split into two sums
\begin{equation}
  \label{eq:qw}
  Q \equiv \sum_{i=0}^{N-1} [E_{x_{i+1}}(\chi_{i+1})-E_{x_{i+1}}(\chi_i)], \quad
  W \equiv \sum_{i=0}^{N-1} [E_{x_{i+1}}(\chi_i)-E_{x_i}(\chi_i)].
\end{equation}
These sums are identified as ``heat'' $Q$ and ``work'' $W$, respectively. In
particular, the meta-work
\begin{equation}
  \label{eq:work}
  W = \sum_{i=0}^{N-1} (x_{i+1}-x_i)\tobs(\chi_i)
\end{equation}
sums the incremental changes of the ``energy'' due to a change of the field
$x$ for the same trajectory.

We can now prove the Jarzynski relation following standard arguments by
combining the form of the path probability Eq.~(\ref{eq:path}) with
Eq.~(\ref{eq:db}). Consider the average
\begin{equation}
  \mean{e^{-W}} = \path{\chi_0\cdots\mathcal{D}\chi_N}
  \rho_{x_0}(\chi_0)p_{x_1}(\chi_0\ra\chi_1)\cdots
  p_{x_N}(\chi_{N-1}\ra\chi_N) e^{-W}
\end{equation}
The first integral reads
\begin{eqnarray}
  \nonumber
\lefteqn{
  \frac{1}{Z(x_0)}\path{\chi_0} \rho_0(\chi_0)p_{x_1}(\chi_0\ra\chi_1)
  e^{-x_1\tobs(\chi_0)} 
  }
  \\ && = \frac{Z_K(x_1)}{Z_K(x_0)}\path{\chi_0}
  \rho_{x_1}(\chi_0)p_{x_1}(\chi_0\ra\chi_1) =
  \frac{Z_K(x_1)}{Z_K(x_0)}\rho_{x_1}(\chi_1).
\end{eqnarray}
Unraveling all terms thus leads to
\begin{equation}
  \label{eq:jarz}
  \mean{e^{-W}} = \frac{Z_K(x_N)}{Z_K(x_0)},
\end{equation}
which is the analogous Jarzynski relation for the meta-work in canonical
ensembles of trajectories.

\subsection{Computing the meta-free-energy $g(x)$}

From Eq.~(\ref{eq:jarz}), we can extract the change of the trajectory (or meta-) free-energy
\begin{equation}
  \Delta g \equiv g(x_N) - g(x_0) = \lim_{K\ra\infty}\frac{1}{K}\ln\mean{e^{-W}}
\end{equation}
from the meta-work. Using this result, the free-energy $g(x)$ of the
$x$-ensemble can be calculated from simulation in the following way. A
trajectory with fixed number of events $K$ is created and equilibrated to the
desired starting value $x_0$ using the $x$-ensemble TPS algorithm described in
appendix A. This is basically a Monte Carlo algorithm accepting or rejecting
proposed trajectories employing the Metropolis criterion. The system then
moves along the ``forward'' path up to the desired maximum value $x_N$ in a
series of steps. For simplicity, we consider a linear protocol
$x_i=x_0+i(x_N-x_0)/N$ although other protocols might be more suitable. Each
step corresponds to a single change to the trajectory whether the proposed
change is accepted under the Metropolis criterion or not.

This process is repeated $M$ times until a good distribution of meta-work for
both the forward and the reverse process (going from $x_N$ to $x_0$) is built
up. The free-energy difference between $x_N$ and $x_0$ can then be computed
with an iterative Bennett's Acceptance Ratio (BAR)
method~\cite{benn76,shir03},
\begin{equation}
  \label{eq:benn}
  \Delta g^{(k+1)} = -\ln\frac{\sum_{j=1}^{M}\left[1+e^{W_{\uparrow,j} -
        \Delta
        g^{(k)}}\right]^{-1}}{\sum_{j=1}^{M}\left[e^{W_{\downarrow,j}}+e^{-\Delta
        g^{(k)}}\right]^{-1}},
\end{equation}
where the sum over $j$ denotes the sum over the \revr{total work}{work values}
for each repetition of forward ($\uparrow$) and reverse ($\downarrow$)
process. \rev{The work values are random numbers with probability
  distributions $P_\uparrow(W)$ and $P_\downarrow(W)$, respectively.}

As is the case in thermodynamic problems, there need be some overlap in the
work distributions for the forward and reverse processes, but the rate at
which these processes occur need not be slow enough to ensure equilibrium at
all points (resulting in completely overlapping work
distributions). 
Strictly speaking, the large-deviation 
function $g(x)$ is defined in the limit of $K \to \infty$.  In practice,
for the numerical estimation of $g(x)$, the length of individual
trajectories as defined by the number of events $K$ is not critical to
the result, provided the meta free-energies are scaled per event. Furthermore,
while short trajectories of low $K$ necessarily require less computation time,
they also necessarily have much larger fluctuations in work distributions,
requiring more repetitions to build a reasonable distribution numerically,
meaning there is some trade off in efficiency.  Note however that a positive aspect of these fluctuations is that the broadening of work distributions can lead to an increase in their overlap.  These considerations indicate that the optimal trajectory
length, and number of steps to calculate the effective meta free-energies as
efficiently as possible, are highly system dependent.


\section{Application to open quantum systems}

For the purpose of demonstrating the validity of the analogous Jarzynski
equality~(\ref{eq:jarz}), we consider simple open quantum systems whose
dynamics are described by Lindblad master equations of the form
\begin{equation}
  \frac{d}{dt} \rho = -i[H,\rho]+\sum_{\alpha = 1}^{N_L} \big (L_\alpha \rho
  L_\alpha ^\dagger - \frac{1}{2} \{ L_\alpha ^\dagger L_\alpha , \rho \} \big),
\end{equation}
where $N_L$ is the number of dissipative terms and the
$L_\alpha$ are the corresponding jump
operators~\cite{lind76,Plenio1998,gard04}. Throughout, $\hbar$ is set to
unity. The countable events are associated with action under the Lindblad
operators (usually photon emission/absorption). Such systems are well suited
to simulation using continuous-time Monte Carlo algorithms \cite{Plenio1998}. 

\subsection{2-Level System}

We consider a laser-driven two-level system, which exchanges photons
with a radiation bath \cite{Plenio1998}. The system is comprised of levels $\ket{0}$ and
$\ket{1}$ with Hamiltonian
\begin{equation}
  H = \Omega(\sigma + \sigma^{\dagger})
\end{equation}
and two jump operators
\begin{equation}
  L_1 = \sqrt{\kappa}\sigma, \qquad L_2 = \sqrt{\gamma}\sigma ^\dagger
\end{equation}
corresponding to photon emission and absorption, respectively. Here $\sigma =
\ket{0} \bra{1}$ and $\sigma^\dagger = \ket{1}\bra{0} $ are lowering and raising operators, and $\Omega$ is the Rabi frequency of the driving laser. As such,
the system emits photo\rev{n}s and is projected 
onto $\ket{0}$ with rate
$\kappa$, and absorbs photons and is projected into $\ket{1}$ state with
rate $\gamma$. The counted events $K$ are any photon emission or absorption,
i.e., the total number of quantum jumps.

We consider first the zero-temperature case ($\gamma=0$), for which there is only one jump - described by action under $L_1$ (photon emission).
The large deviation function in this case reads 
\begin{equation}
g(x)=-3\ln{\left(1+ \frac{x}{2}\right)} 
\label{gx}
\end{equation}
This result is obtained by inverting
$g(x)=\theta^{-1}(x)$, where $\theta(s)$ is the largest eigenvalue of the
deformed master equation associated with the $s$-ensemble, see
Ref.~\cite{budi14} for details. 

\begin{figure}[t]
  \centering
  \includegraphics{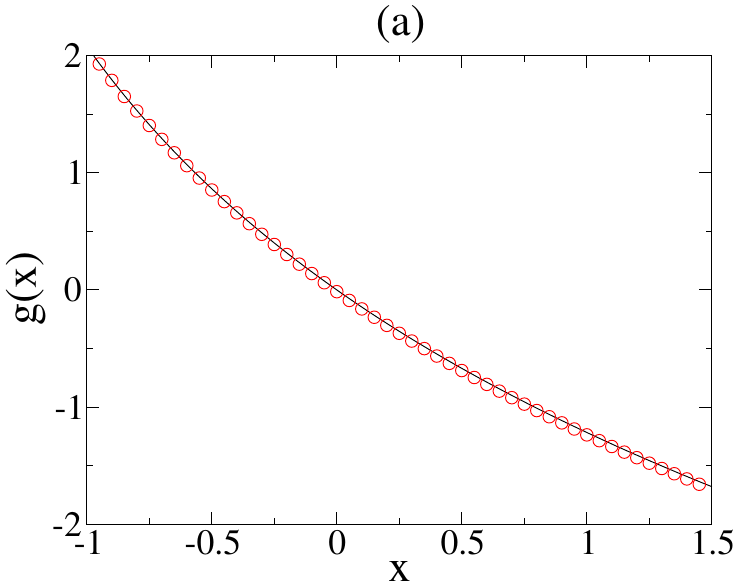}\includegraphics{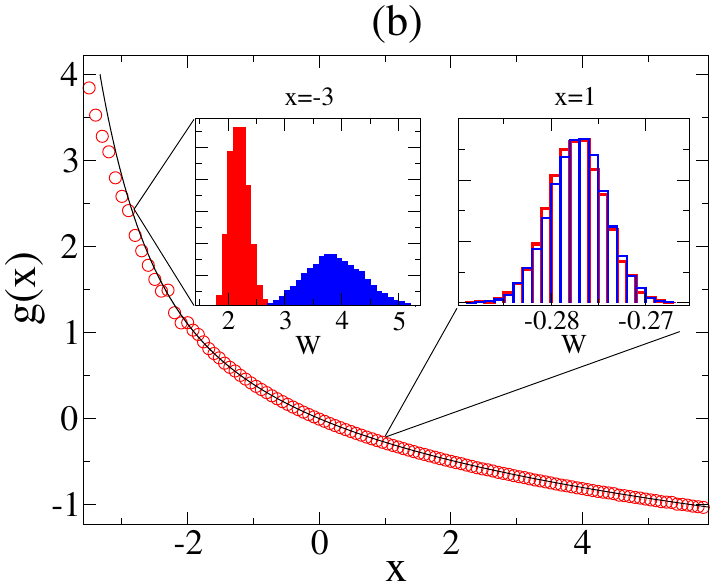} 
  \caption{2-level system.  (a) Comparison of the meta-free-energy $g(x)$
    obtained numerically via the trajectory Jarzynski relation (symbols) to
    the exact analytical result (\ref{gx}) (solid line) for a range of $x$, in
    the zero temperature case, with $\kappa=4\Omega$.  (b) Same as in (a), but
    now for the finite temperature case, with $\kappa=6\Omega$ and
    $\gamma=2\Omega$. The statistical error is smaller than the symbol sizes.
    Insets to (b): \rev{Sampled histograms for the} meta-work
    distribution\revd{s} $P_\uparrow(W)$ for the forward\revd{s} (red) and
    $P_\downarrow(-W)$ for the backward\revd{s} process (blue), at the two
    \rev{final} values of $x$ shown.}
  \label{fig:f0}
\end{figure}

Figure ~\ref{fig:f0} provides a numerical test of the Jarzynski relation (\ref{eq:jarz}) for trajectories with $K=20$ events. We have sampled $M=5000$ trajectories for the forward and backward
protocol, where trajectories started from an initial $x_0=0$ (equilibrium)
state to a final state ranging between $x=-1$ and $x=1.5$, with $N=1000$ TPS
step moves for each direction. As criterion to stop the BAR iterations, we
chose the threshold $10^{-5}$ for the fractional change of the estimated
$g(x)$ between iterations. For this system convergence is reached very fast
taking typically 2-3 iterations, and there is a good agreement between the results obtained from the Jarzynski relation and the exact results.

We now consider the finite temperature case with parameters $\kappa = 6\Omega$, $\gamma = 2\Omega$. Here action under both $L_1$ and $L_2$ occurs, and so there are two jump possibilities. Fig.~\ref{fig:f0}(b) provides a numerical test of the Jarzynski relation in this case. Analytical results are again obtained from the largest
eigenvalue of the deformed master operator corresponding to the
$s$-ensemble, and inverted to give the $x$-ensemble meta-free-energy $g(x)$. The exact expression is available but cumbersome and rather
unilluminating to be given explicitly. Note that the true $g(x)$ diverges
close to $x\simeq-3.5$ [cf.\ with the zero temperature case, Eq.\ (\ref{gx}), where the limiting value is $x=-2$]. Again, $M=5000$ iterations were used for trajectories
of $K=20$ events but with now $N=5\times 10^5$ TPS step moves for each
iteration. While there is a good agreement between the results obtained from the Jarzynski relation and the exact results for a broad range of $x_N$, we have extended the plotted range of $x$ values to
demonstrate that the numerical estimate for $g(x)$ starts to divert from its
analytical prediction as we approach the divergence. For $x<0$ the
``pressure'' is negative, selecting rare trajectories with large trajectory
length $\tobs$. Our numerical procedure breaks down because it takes an
increasing amount of time to equilibrate the system at the final $x$ for the
backward iterations. For the forward-backward protocol, $N$ has to be
sufficiently large to generate work distributions that sufficiently overlap in
order for Eq.~(\ref{eq:benn}) to work. This is demonstrated in the inset of
Fig.~\ref{fig:f0}(b). This is a
general feature of the Jarzynski relation. Although in principle it holds for
any driving speed and any protocol, application to data requires either to
sample extreme work values sufficiently or to generate distributions from
forward and backward protocols that overlap.

\subsection{Micromaser}

The micromaser provides a useful test of a pseudo-many-body system, as well as
a system with many first-order phase transitions in the $x$-ensemble. A
detailed account of the model can be found in Ref.~\cite{engl02} and is only
briefly summarized here.  A cavity is pumped by excited two-level atoms and it
also interacts with a thermal bath. The system is described by \rev{a} single
bosonic mode evolving according to a Lindblad master equation with four
Lindblad terms, two corresponding to the cavity-atom interactions,
\begin{equation}
  L_1 = \sqrt{r} \frac{\sin
    (\phi \sqrt{aa^\dagger})}{\sqrt{aa^\dagger}}, \qquad
  L_2 = \sqrt{r} \cos (\phi\sqrt{aa^\dagger}),
  \label{L12}
\end{equation}
and two corresponding to the cavity exchanging photons with a radiation bath,
\begin{equation}
  L_3 = \sqrt{\kappa}, \qquad
  L_4 = \sqrt{\gamma} a^\dagger.
  \label{L34}
\end{equation}
Here, $a$ and $a^\dagger$ are the raising and lowering operators of the
cavity mode, respectively, and $\kappa$ and $\gamma$ are the rates of photon
emission and absorption to/from the radiation bath. The parameter $\varphi$
encodes the information on the atom-cavity interaction and $r$ is the atom
beam rate through the cavity. The system can be parametrised by a single
``pump parameter'' $\alpha\equiv\varphi\sqrt{r/(\kappa-\gamma)}$. The events being counted are the actions under any of the four Lindblad
terms.

\begin{figure}[t]
  \centering
  \includegraphics[width=.5\linewidth]{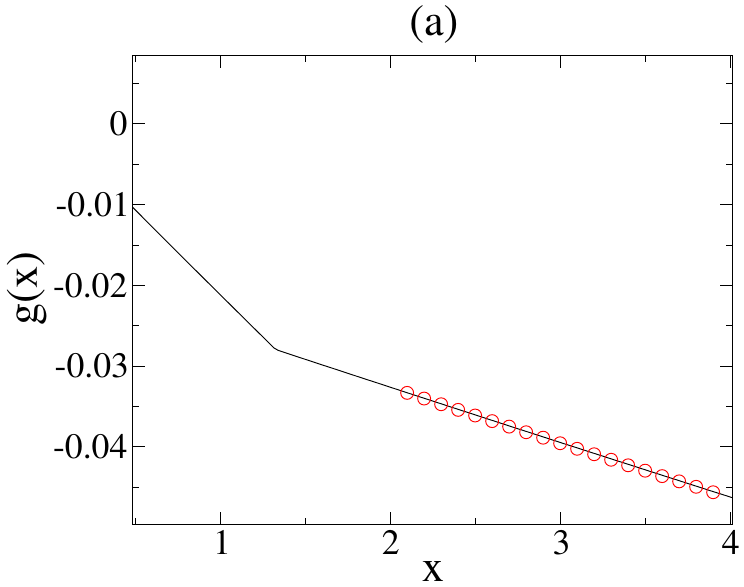}\includegraphics[width=.5\linewidth]{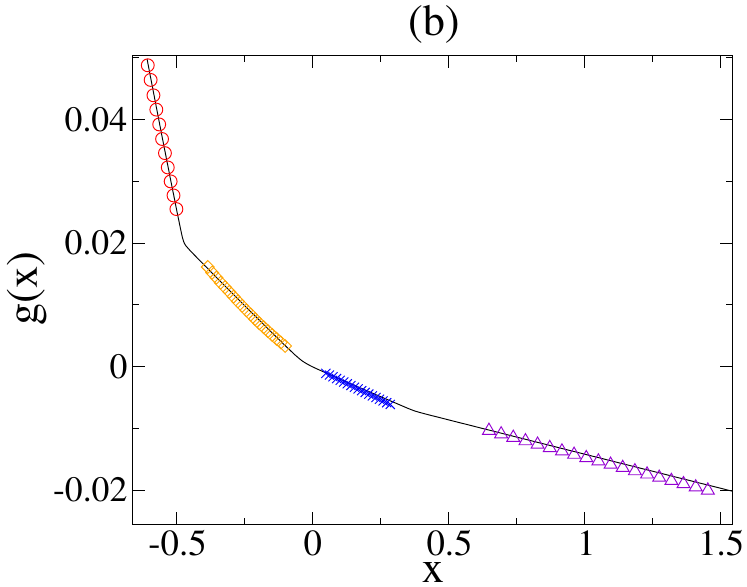}
  \caption{ Micromaser. (a) Comparison of the meta-free-energy $g(x)$ obtained numerically via the trajectory Jarzynski relation (symbols) to results obtained by direct diagonalisation of the master operator (solid line) for $\alpha =1.2\pi$, where the system is initially equilibrated to $x=2$. (b) Same as in (a), but now for $\alpha = 4\pi$. Different simulations, equilibrated to different initial values of $x$ are denoted by different symbols.}
  \label{fig:f3}
\end{figure}

Despite being a system with a single degree of freedom, the micromaser has a rich dynamical behaviour due to the combination of an infinite dimensional Hilbert space and the non-linear jump operators $L_1$ and $L_2$.  In particular, it displays a number of distinct dynamical phases and transitions between them \cite{garr11,Horssen2012,Lesanovsky2013}. (Strictly speaking, these are sharp crossovers which only become singular in the limit of $r \to \infty$; see \cite{Catana2012,Horssen2012}.)  Note also that in the dynamics generated by the operators (\ref{L12}-\ref{L34}) an initial density matrix that is diagonal stays diagonal for all times.  Due to this, the dynamics of the micromaser, while quantum in origin, is in effect that of a classical stochastic system.

We first attempt to compute meta-free-energy differences within a single phase. Fig.~\ref{fig:f3}(a) provides a numerical test of the Jarzynski relation for a pump parameter of $\alpha=1.2\pi$. The trajectories are initially equilibrated to a non-equilibrium dynamical phase with $x_0 =2$, and the Jarzynski protocol run for trajectories of $K=1000$ jumps, with $M=5000$ iterations and $N=60000$ TPS step moves per iteration. The computed meta-free-energy differences are compared to results obtained from direct diagonalisation of the master operator, as in \cite{garr11}, and a good agreement is found between the two methods.
Provided the existence of phases, and the boundaries between them, is known, a complete picture of meta free-energy differences can be constructed even when there are multiple dynamical phases. For example, with the pump parameter taking a value of $\alpha = 4\pi$, four distinct phases occur \cite{garr11,Horssen2012}, see Fig.~\ref{fig:f3}(b), and $g(x)$ can be computed within phases by initially equilibrating the trajectories to a value of $x$ within the required phase. Again trajectories of $K=1000$ jumps were used, with $M=5000$ iterations and $N=60000$ TPS step moves per iteration.

\subsection{Driving across a first-order phase transition}

We finally examine the behavior of the Jarzynski relation using a protocol
$x_0\ra x_N$ that crosses a phase boundary $x^\ast$ at a finite speed. In the
quasi-stationary limit of $N\ra\infty$, we obtain from the definition
Eq.~(\ref{eq:Z}) the well-known expression
\begin{equation}
  \label{eq:ti}
  \ln\frac{Z_K(x_N)}{Z_K(x_0)} = \IInt{x}{x_0}{x_N} \pd{\ln Z_K(x)}{x} =
  -\IInt{x}{x_0}{x_N} \mean{\tobs}_x
\end{equation}
for thermodynamic integration, where the subscript emphasizes that the average
is calculated from equilibrated trajectories at fixed $x$. Eq.~(\ref{eq:ti})
is known to fail in the presence of a discontinuous phase transition, not
because the equation is wrong but because of the way a simulation is carried
out in practice. Typically, one will apply a small change $x_i\ra x_{i+1}$,
let the system relax, and then record data to calculate the average. Crossing
$x^\ast$, the system will not immediately adapt to the new state but follow
the metastable branch due to the cost of nucleating the new stable phase,
thus violating the assumption that the calculated mean corresponds to the true
equilibrium mean.  In the micromaser, sharp crossovers occur at certain values of the biasing field between phases that can be characterised by either their average emission rate, or the closely related expected photon occupation of the cavity~\cite{garr11,Horssen2012,Lesanovsky2013}.  
When considering these transitions in the context of the $x$-ensemble,
different phases have significantly different average trajectory lengths for
the same fixed number of quantum jumps \cite{budi14}.  Just like in ordinary
first-order transitions, pronounced metastability may prevent from estimating
meta free-energies accurately with (\ref{eq:ti}). 
This can occur when the transition at $x^\ast$ is between phases with very different activities.  In this case, if trajectories are prepared in the less active phase (for example starting from $x=0$ and increasing $x$), the barrier to nucleate the more active phase when $x>x^\ast$ can be prohibitive for practical simulation.  The nucleation event can be promoted externally, for example by altering the photon occupation of the cavity by temporarily increasing the pump parameter (or similar ``parallel tempering''). 
But without such external interference the timescale for nucleating the new stable phase is often beyond what can be reasonably simulated.

\begin{figure}[t]
  \centering
  \includegraphics{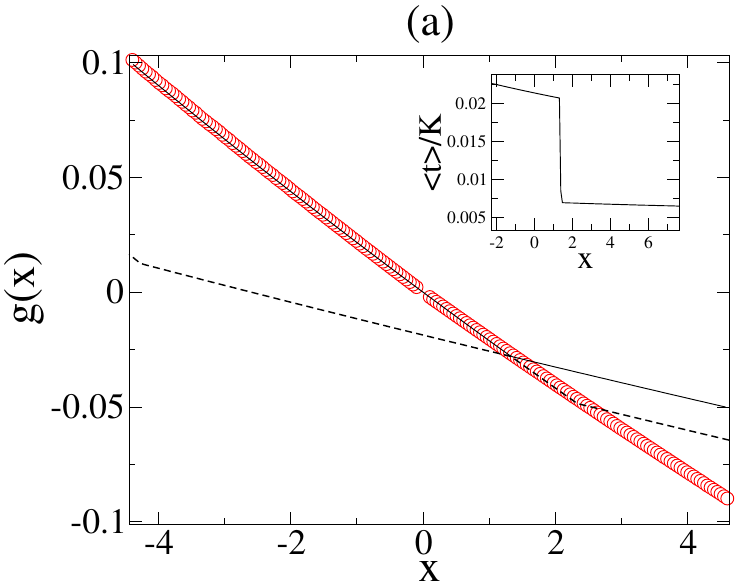} \includegraphics{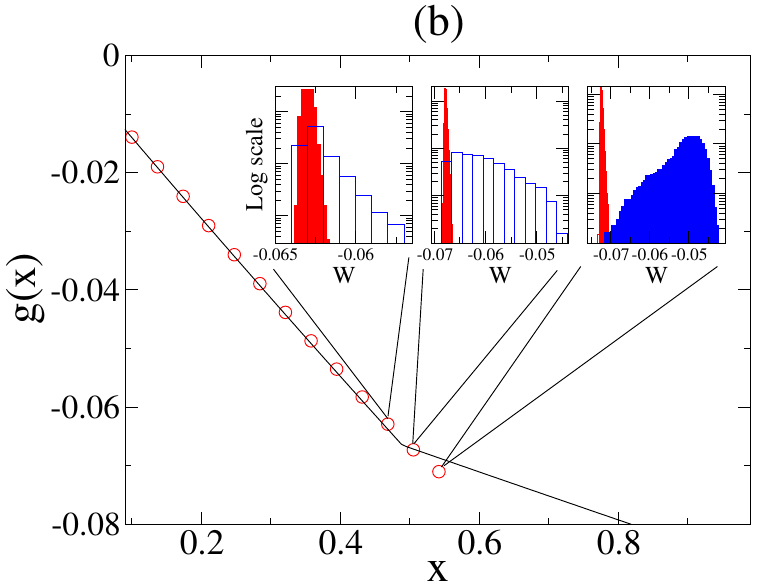} 
  \caption{Micromaser with cross-phase Jarzynski protocol. (a) Comparison of
    the numerical meta free-energy, $g(x)$, obtained numerically via the
    Jarzynski relation (symbols) to results obtained by direct diagonalisation
    (solid line) in a micromaser with pump parameter $\alpha = 1.2\pi$, at a
    finite temperature (${\gamma}/{\kappa}=0.15$). The second largest
    eigenvalue (dashed line) is plotted to illustrate the meta free-energy
    calculation being locked to the metastable branch after the
    transition. Inset to (a): the expected waiting time per event showing the
    differing dynamic properties of the two phases. (b) Same as in (a), but
    now at zero temperature (${\gamma}/{\kappa}=0$). Insets to (b):
    \rev{sampled} meta-work distributions for the forward\revd{s} (red) and
    backward\revd{s} (blue) process for the three points shown.}
  \label{fig:f4}
\end{figure}

One could hope that the Jarzynski relation, given that it applies to arbitrarily fast non-equilibrium protocols, would provide a way out of this problem
since trajectories can be sampled at finite rate for the change in $x$.  In practice, however, even with slow driving speeds it is problematic to compute free-energy differences across first-order phase boundaries.  Results for the micromaser are
shown in Fig.~\ref{fig:f4} (for a pump parameter of $\alpha=1.2\pi$ and with
$\frac{\gamma}{\kappa}=0.15$ corresponding to a temperature $T=0.5$ \cite{garr11,budi14}).  Trajectories with $K=2000$ jumps were
sampled for $M=5000$ iterations, with $N=10^6$ TPS step moves for each
iteration. For the chosen parameters, the system is known to undergo a first-order transition
at $x^\ast\simeq 1.34$ \cite{budi14}. The computed free-energy difference using the Jarzinsky relation gets locked to the phase that is stable for $x<x^\ast$ but which becomes metastable for $x>x^\ast$.  This is evident by the fact that the computed free-energy follows the path of the eigenvalue that dominates for $x<x^\ast$, but which becomes subdominant at $x>x^\ast$.

This locking in the metastable branch occurs even if one reduces the  difference in the dynamic properties between the two phases, for example by considering zero temperature and for smaller beam rate \cite{garr11}, or improves the sampling (for example by doubling the number of interations), see Fig.\ \ref{fig:f4}(b).  The cause can be understood by looking at the meta-work distributions for the forward and reverse processes,  see insets to Fig.\ \ref{fig:f4}(b).  For the conditions shown, the driving is slow enough for the forward and reverse meta-work distributions to overlap immediately before the phase transition.  However as the phase boundary is crossed the two become separated. A small residual spike of the reverse distribution lies within the bulk of the forward distribution, corresponding to a small fraction of cases where the reverse process starts in the metastable phase. This occurs precisely because the simulation cannot be done in the ``thermodynamic limit'' of $K\rightarrow \infty$ and $r/(\kappa - \gamma) \rightarrow \infty$, i.e.\, the transition is not strictly a phase transition but a very sharp crossover \cite{Horssen2012}.  Thus when differences in the meta free-energy $g(x)$ is computed with the BAR method, it only sees the metastable phase.  It is worth noting that these attempts to compute a cross-phase free-energy difference took two orders of magnitude more computation than any of the single-phase free-energy computations.

\section{Concluding remarks}

We have extended the ``thermodynamics of trajectories'' method to show the
existence of analogous fluctuation theorems associated with corresponding
``non-equilibrium'' processes.  In particular, we have studied the analogous
Jarzynski relation connecting meta-work to changes in trajectory
free-energies.  For convenience, we have considered ensembles of trajectories
characterised by a fixed number of configuration changes (or jumps) and
variable overall time \cite{budi14}.  The parameter that was driven was the
field conjugate to the total trajectory time, and the meta-time associated to
this non-equilibrium procedure was that of the TPS scheme used to evolve
trajectories in trajectory space.  The associated work, or meta-work, was
given by the path integral of the change in average total trajectory time,
i.e., the change in the trajectory observable conjugate to the driven field,
again in analogy with what occurs with configuration ensembles.  The analogous
Jarzynski relation connects the average of the exponential of this meta-work
over the driven process to the difference of the large-deviation rate
functions that determine the trajectory ensembles at the endpoint values of
the driven field.  Similar relations hold in other trajectory ensembles, for
example that of trajectories of fixed total time and where the number of
events fluctuates.

Our results here further underpin the thermodynamics approach to dynamics.
Not only ensembles of dynamical trajectories can be studied by generalising
equilibrium statistical mechanics via large deviation methods, but also
non-equilibrium statistical mechanics tools can be generalised and applied to
uncover properties of such ensembles.  By considering the analogous Jarzynski
relation we have shown that the large-deviation function that encodes the
properties of one trajectory ensemble can be obtained \revd{from the one from
  another} by considering the statistics of the meta-work performed as the
parameter that characterises the ensemble\revd{s} is driven.

A further interesting observation is the following.  The general relation
between forward and backward processes that underpins most integral
fluctuation theorems is a straightforward consequence of probability
conservation \cite{Seifert2012}.  \revr{Very f}{F}ew integral fluctuation
relations are ``non-trivial'' in the sense of conveying actually useful
information about the problem studied.  This occurs when one can write the
stationary distribution in terms of ``weights'' that encode their functional
dependence on the objects that form the ensemble under consideration (usually
configurations; trajectories in our case), and a ``free-energy''.  For
ensembles of configurations, \revr{we can think of three such cases: that of
  the canonical equilibrium ensemble, leading to}{these include} the Jarzynski
relation proper \cite{jarz97,jarz97a}\revr{; that of driven stationary states
  leading to}{ and} the Hatano-Sasa relation~\cite{Hatano2001} \rev{for driven
  stationary states.}\revd{; and *** Seifert relation ***.} We note that the
class of trajectory ensemble problems we studied here adds to this small
group.  These are cases where the ``normalisation constant'' of the stationary
probability distribution also has physical meaning, as it is given by the
large-deviation function which is the generating function for moments and
cumulants of time-integrated and thus play the role of trajectory
free-energies.

\section*{Acknowledgments}
This work was supported by Leverhulme Trust Grant No. F/00114/BG.

\appendix
\section{Sampling algorithm}
\label{sec:algo}

For completeness, here we describe the algorithm used to sample
trajectories. This algorithm is an adaptation of the Crooks-Chandler
method~\cite{croo01} described in section 3.4 of Ref.~\cite{dell02}; see also \cite{Bolhuis2008,budi14}.

\begin{enumerate}
\item Fix total event numbers $K$
\item Generate and store a random number/set of random numbers, $\{r\}_i$ as needed to describe each event, defining a complete trajectory, $\chi$.
\item Calculate the total time taken by the trajectory, $\tobs$.
\item Set $x$ to 0.
\item Randomly select and modify a single random number set, $\{r\}_i\to\{r'\}_i $ to propose a new trajectory, $\chi'$
\item Recalculate the event $\{r\}_i$, and any subsequent events that are altered by the modified result of event $i$. If at any point the state of trajectory $\chi'$ is identical to that of $\chi$ after jump $i+\Delta i$ further computation of the trajectory is unnecessary.
\item Calculate the new trajectory length, $\tobs'$
\item Accept/Reject the new trajectory based on the metropolis acceptance critera $P_\mathrm{accept} = \min\{ 1,e^{-x(\tobs'-\tobs)}\}$
\item Repeat steps (v)-(viii) until the trajectory is equilibrated to the current values of $x$
\item Increment $x$ by some small $\delta x$
\item Repeat steps (v)-(xi) until the desired final value of $x$ is reached
\end{enumerate}


\section*{References}

\bibliographystyle{prsty}
\bibliography{refs,refsv2}

\end{document}